\documentclass[preprint,showpacs,preprintnumbers,amsmath,amssymb]{revtex4}

\usepackage{graphicx}
\usepackage{dcolumn}
\usepackage{bm}

\begin{document}

\vspace{5mm}

\newcommand{\goo}{\,\raisebox{-.5ex}{$\stackrel{>}{\scriptstyle\sim}$}\,}
\newcommand{\loo}{\,\raisebox{-.5ex}{$\stackrel{<}{\scriptstyle\sim}$}\,}

\title{Formation of hypernuclei in evaporation and fission processes}

\author{A.S.~Botvina$^{1,2}$, N.~Buyukcizmeci$^{3}$, 
A.~Ergun$^{3}$, R.~Ogul$^{3}$, M.~Bleicher$^{1}$, J.~Pochodzalla$^{4}$}

\affiliation{$^1$Frankfurt Institute for Advanced Studies and ITP J.W. Goethe 
University, D-60438 Frankfurt am Main, Germany} 
\affiliation{$^2$Institute for Nuclear 
Research, Russian Academy of Sciences, 117312 Moscow, Russia} 
\affiliation{$^3$Department of Physics, Selcuk University, 42079 Kampus, 
Konya, Turkey}
\affiliation{$^4$ Helmholtz-Institut Mainz and Institut f{\"u}r Kernphysik, 
J.Gutenberg-Universit{\"a}t Mainz, D-55099 Germany}

\date{\today}

\begin{abstract}

There are excellent opportunities to produce excited heavy hyper residues 
in relativistic hadron and peripheral heavy-ion collisions. We investigate 
the disintegration of such residues into hyper nuclei via evaporation of 
baryons and light clusters and their fission. Previously these processes 
were well known for normal nuclei as the decay channels at low excitation 
energies. We have generalized these models for the case of hyper-matter. 
In this way we make extension of nuclear reaction studies at low temperature 
into the strange sector. We demonstrate how the new decay channels can be 
integrated in the whole disintegration process. Their importance for 
mass and isotope distributions of produced hyper-fragments is emphasized. 
New and exotic isotopes obtained within these processes may provide a unique 
opportunity for investigating hyperon interaction in nuclear matter. 

\end{abstract}


\maketitle

\section{Introduction}

Hypernuclei are formed when hyperons ($Y=\Lambda,\Sigma,\Xi,\Omega$) 
produced in high-energy interactions are captured by nuclei. They live 
significantly longer than the typical reaction times. 
Baryons with strangeness embedded in the nuclear environment provide the 
only available possibility to approach the many-body aspect of the strong 
three-flavor interaction at low energies. In the same time, hypernuclei 
can serve as a  tool to study the hyperon--nucleon and hyperon--hyperon 
interactions. The investigation of reactions leading to hypernuclei and 
the structure of hypernuclei is the progressing 
field of nuclear physics, because it provides complementary methods to 
improve traditional nuclear studies and open new horizons for studying 
particle physics and nuclear astrophysics 
(see, e.g., \cite{Ban90,Sch93,Has06,Gal12,Buy13,Hel14} 
and references therein). 

Traditionaly for the hypernuclear physics focuses on 
spectroscopic information and is dominated by a quite limited set of 
lepton- and hadron-induced reactions \cite{Ban90,Has06}. In these reactions 
the directly produced kaons are often used for tagging the production of 
hypernuclei in their ground and low excited states. However, 
very encouraging results on hypernuclei were obtained in experiments with 
relativistic ion collisions \cite{star,alice,saito-new} and in other 
reactions where a large amount of energy is deposited in nuclei 
\cite{Arm93,Ohm97}. 
Many experimental collaborations PANDA \cite{panda}, CBM \cite{Vas15}, 
HypHI, Super-FRS, R3B at GSI Facility for Antiproton and Ion reserach (FAIR) \cite{super-frs}, BM@N and MPD 
at Nuclotron-based Ion Collider Facility (NICA) \cite{nica}) plan to investigate hypernuclei and 
their properties in reactions induced by relativistic hadrons and ions. 
The limits in isospin space, particle unstable states, multiple strange 
nuclei, and precision lifetime measurements are unique topics of these 
fragmentation reactions. 

We especially emphasize a possibility to form hypernuclei in the 
deep-inelastic reactions leading to fragmentation processes, as they 
were discovered long ago \cite{Dan53}. 
As already discussed \cite{Buy13,Bot11}, in these reactions initiated by 
high-energy hadrons, leptons, and ions one can get a very broad distribution 
of produced hypernuclei including the exotic ones and with the extreme 
isospin. This can help to investigate the structure of nuclei by extending 
the nuclear chart into the strangeness sector \cite{Ban90,Sch93,Has06}. 
Complex multi hypernuclear systems incorporating more than 
two hyperons can be created in the energetic nucleus-nucleus collisions, 
and this may be the only conceivable method to go even beyond $|s|=2$.
An essential theoretical progress was achieved in the investigation of the 
normal nuclear reactions associated with both peripheral relativistic 
heavy-ion collisions and hadron-induced reactions (see, e.g., 
\cite{Bon95,Xi97,Sch01,Ogu11} and references therein). This gives us an 
opportunity to apply well known theoretical methods adopted for the description 
of these reactions also to the production of hypernuclei \cite{Bot07,Das09}. 
In this paper we generalize the two very popular nuclear reaction models, 
the evaporation of light particles from the excited compound nucleus and 
the fission of the compound nucleus, for the description of the decay 
of excited hypernuclei.
We investigate an important case of low excitation energies, 
in addition to high excitations leading to multifragmentation processes, 
which were analyzed previously in Refs.~\cite{Bot07,Buy13}. As we show, 
many novel possibilities arise for formation of hypernuclei. New 
experiments on hypernuclei, in particular at GSI/FAIR and other 
accelerators, may be directed by employing such production mechanisms.

\section{Production of excited hyper-residues}

The hyperons are produced in high-energy particle reactions, 
e.g., nucleus-nucleus, hadron-nucleus and lepton-nucleus collisions. 
Usually, the emission of many particles accompanies the production of 
hyperons and an initial nucleus can lose many nucleons. As known from 
the interactions in normal nuclei, these processes will lead to a broad 
spectrum of excitations of remaining residual nuclei \cite{Xi97,Ogu11}. 
For this reason the 
possibility to capture a hyperon will be realized mostly at an excited 
nucleus. We should note that a direct hyperon 
capture in the nuclear ground state has a very 
small probability. It is an important advantage of deep-inelastic processes 
that they allow for forming hyper-residues with very broad distribution 
in mass and excitation energy \cite{Bot11,BotG16}. 

The modifications in normal nuclei after the interaction with high energy 
hadrons and leptons are very well described in the literature (see, e.g., 
Ref.~\cite{Bon95} and references therein). The production of strangeness is one 
of the possible channels and, besides the capture of few hyperons, we do 
not expect an essential change in the structure of a residue in this 
case. In comparison with hadron-induced reaction, 
the peripheral relativistic heavy-ion collisions lead to the 
larger number of individual nucleon-nucleon interactions and, as a result, 
to a larger number of produced particles related to a larger loss of nucleons from 
the residues. However, we have the same qualitative picture of what happens 
in the residues. As an example we refer to ion collisions in the following. 
It was demonstrated in the previous works \cite{Bot13,Bot15} that the yields 
of the hypernuclear residues in peripheral ion collisions will saturate with 
energies above 3--5 $A$ GeV (in the laboratory frame). Therefore, the 
accelerators of moderate relativistic energies can be used for the intensive 
studies of hypernuclei. The subthreshold production of hyperons 
becomes possible in these reactions down to the energies of $\sim$1 $A$ GeV 
\cite{BotG16}. 
At the laboratory energies of ions around 1--2 $A$ GeV one can effectively 
obtain nuclei with the modern experimental fragment separation 
methods \cite{Rap13a,aumann,frs}. This gives chances to measure many new 
exotic hypernuclei. Another research direction is related to increasing the 
energy up to $\sim$10  $A$ GeV, when there is an opportunity to produce 
multistrange hyper-fragments which can be measured with the high precision 
detectors, for example, by CBM collaboration at FAIR \cite{Vas15}. 

The typical excitation energies of the residues can be found from analysis 
of fragmentation/multifragmentation experimental data \cite{Bon95,Xi97,Ogu11}, 
as well as from the model calculations \cite{BotG16}. Both ways are 
consistent, and we have obtained the excitation energies from 0 to around 8 
MeV per nucleon for these residues. The upper limit is naturally consistent 
with the nuclear binding energy, where the nuclei can still live for a time 
($\sim$100 fm/$c$) sufficient for development of the collective decay modes (as 
multifragmentation).  
Because the hyperon life time in nuclei is essentially larger than the time 
for decay of excited nuclei, we must consider the de excitation processes 
leading to the production of really cold hypernuclei. As demonstrated 
previously, these processes are very promising for obtaining novel 
hypernuclei in the case of the multifragmentation break up at high 
excitation energy \cite{Bot07,Buy13}. In this paper we 
investigate the region of low excitation energies, where evaporation and 
fission decay modes dominate. 
Such excitations may be obtained in not-very-complicated reactions 
involving only few nucleons. For example, one nucleon can interact with an 
incident particle and be knocked out from a heavy nucleus. In addition, 
$K^{+}$ and a low-energy $\Lambda-$ hyperon may be produced. In the following 
$K^{+}$ can escape, and $\Lambda$ may be captured inside the nucleus. 
In this case the background for hypernuclear experimental measurements is 
minimal. The "hole" in a nucleus from the nucleon can contribute with around 
20 MeV (on average) to the nucleus excitation, whereas the hyperon capture 
may add another 10--20 MeV. In reality, however, the interactions with other 
nucleons may lead to more higher excitation energies. To describe the 
de excitation 
of low-excited hyper nuclei, we generalize the corresponding nuclear 
evaporation and fission models. We believe this generalization is possible 
because 
the hyperon-nucleon interaction is of the same order as the nucleon-nucleon 
one, and the hyperon potential in 
a nucleus is considered as around 2/3 of the nucleon potential.

\section{De-excitation  of hyper-nuclei}

For completeness, we provide knowledge about all main secondary 
de excitation processes, because they are complementary to each other. We expect 
the existence of hypernuclear decay mechanisms which are similar to the 
decay of normal nuclei. 

\subsection{Decay of light hyper-nuclei}

We remind the reader that in the case of interaction with light nuclei ($A\loo 12-16$) 
the excited light hyper residues are produced after the dynamical stage. 
For their disintegration one can use the Fermi-break up model 
 \cite{Botvina87,Bon95} generalized 
by including $\Lambda$ hyperons in Ref.~\cite{lorente}. In the microcanonical 
approximation we take into account all possible break up channels, 
which satisfy the mass number, hyperon number (i.e., strangeness), charge, 
energy, and momentum conservation and simulate the competition between 
these channels. The probability of each break up channel ch is 
proportional to the occupied phase space and the 
statistical weight of the channel containing $n$ particles with 
masses $m_{i}$ ($i=1,\cdots,n$) can be calculated as 
\begin{eqnarray}
\label{eq:Fer} W_{ch}^{mic}\propto
\frac{S}{G}\left(\frac{V_{f}}{(2\pi\hbar)^{3}}\right)^{n-1}
\left(\frac{\prod_{i=1}^{n}m_{i}}{m_{0}}\right)^{3/2}
\nonumber\\
~\cdot \frac{(2\pi)^
{\frac{3}{2}(n-1)}}{\Gamma(\frac{3}{2}(n-1))}\cdot
\left(E_{kin}-U_{ch}^{C}\right)^{\frac{3}{2}n-\frac{5}{2}},
\end{eqnarray}
where $m_{0}=\sum_{i=1}^{n}m_{i}$ is the summed mass of the
particles, $S=\prod_{i=1}^{n}(2s_{i}+1)$ is the spin degeneracy
factor ($s_{i}$ is the $i$-th particle spin),
$G=\prod_{j=1}^{k}n_{j}!$ is the particle identity factor ($n_{j}$
is the number of particles of kind $j$). $E_{kin}$ is the kinetic 
energy of nuclei and $U_{ch}^{C}$ is the Coulomb interaction energy 
between nuclei, which are related to the energy balance 
as described in Ref.~\cite{lorente}. 
The table masses of both ground states and known excited states of 
(hyper-)nuclei (see, e.g., Refs. \cite{Ban90,Has06}) are included. 
We have obtained in this case very encouraging predictions on the 
hypernuclei production  \cite{lorente,Bot13}. 

\subsection{Sequential decay models: evaporation and fission}

The successive particle emission from large hot primary nuclei 
is one of the basic de-excitation mechanism and it was implemented 
for the decay of normal compound nuclei nearly 60 yr ago 
\cite{Weis37}. This mechanism has been under intensive theoretical study 
and it is realized in 
many versions which provide very good descriptions of experimental data 
(e.g., see discussion in Ref.~\cite{Bon95}). In this work we consider 
the generalization of the evaporation developed in 
Refs.~\cite{Botvina87,Bon95,Buy05,Hen10,Ima15}, and extend it for hyper 
matter. For excited hyper-nuclei the 
modification of the standard evaporation scheme is the following: 
Besides emission of normal light particles (nucleons, $d$, $t$, $\alpha$, 
and others up to oxygen) in ground and particle-stable excited states 
\cite{Botvina87}, we take into account the emission of strange particles 
($\Lambda$-hyperon, $^{3}_{\Lambda}$H, $^{4}_{\Lambda}$H, $^{4}_{\Lambda}$He, 
$^{5}_{\Lambda}$He, and $^{6}_{\Lambda}$He). The width for the emission 
of a particle $j$ from the  compound  nucleus $(A,Z)$ is given by
\begin{equation} \label{eq:eva}
\Gamma_{j}=\sum_{i=1}^{n}\int_{0}^{E_{AZ}^{*}-B_{j}-\epsilon_{j}^{(i)}}
\frac{\mu_{j}g_{j}^{(i)}}{\pi^{2}\hbar^{3}}\sigma_{j}(E)
\frac{\rho_{A^{'}Z^{'}}(E_{AZ}^{*}-B_{j}-E)}{\rho_{AZ}(E_{AZ}^{*})}EdE.
\end{equation}
Here the sum is taken over the ground and all particle-stable excited states 
$\epsilon_{j}^{(i)}~(i=0,1,\cdots,n)$ of the fragment $j$, 
$g_{j}^{(i)}=(2s_{j}^{(i)}+1)$   is  the 
spin degeneracy factor of the $i$th excited  state, 
$\mu_{j}$ and $B_{j}$ are corresponding reduced mass and separation energy, 
$E_{AZ}^{*}$ is the excitation energy of the initial (mother) nucleus, and $E$ is the kinetic energy of an emitted particle in the center-of-mass 
frame. In Eq. (\ref{eq:eva}) $\rho_{AZ}$ and $\rho_{A^{'}Z^{'}}$ are 
the level densities of the initial $(A,Z)$ and 
final (daughter) $(A^{'},Z^{'})$ compound  nuclei in the evaporation chain.  
The cross section $\sigma_{j}(E)$ of the inverse 
reaction $(A^{'},Z^{'})+j=(A,Z)$ was calculated using the optical model 
with nucleus-nucleus potential \cite{Botvina87}. This evaporation 
process was simulated by the Monte Carlo method and the conservation of 
energy and momentum was strictly controlled in each emission step. 
After the analysis of experimental data we come to conclusion that at 
sufficient large excitation energies (more than 1 MeV per nucleon) it is 
reasonable to include the decreasing symmetry energy coefficient in 
mass formulas, which leads to adequate description of isotope distributions 
\cite{Buy05,Hen10,Ima15}. 

By considering the de excitation of hypernuclei 
we have taken into account their hyper energy term. In 
consistence with our previous works we suggest to use a reliable mass 
formula introduced in Ref.~\cite{Bot07,Buy13}, where the binding hyper energy 
$E_{b}^{\rm hyp}(A,H)$ is parametrized as 
\begin{equation} \label{hyp}
E_{b}^{\rm hyp}(A,H)=(H/A)\cdot(10.68 A - 21.27 A^{2/3}) MeV .
\end{equation}
In this formula $H$ is the hyperon number and 
the binding energy is proportional to the fraction
of hyperons in the system ($H/A$). The second part represents the
volume contribution reduced by the surface term and thus resembles the 
liquid-drop parametrization based on the saturation of the nuclear
interaction. As demonstrated in Ref.~\cite{Buy13} the formula gives 
a reasonable description of binding energies of known hypernuclei. 
A captured $\Lambda$-hyperon can occupy the s-state deep inside nuclei, 
because it is not forbidden by the  Pauli principle. For this reason adding 
this hyperon to nuclei is a more effective way to increase their binding 
than adding nucleons, especially for large species. We apply the 
same formulas (\ref{eq:eva}) for emission of hyperons and light hypernuclei, 
however, by taking into account that the additional hyper terms 
must appear in the corresponding separation energies $B_{j}$,
\begin{equation} \label{sep}
\Delta B_{j} = E_{b}^{\rm hyp}(A,H) -  
E_{b}^{\rm hyp}(A^{'},H^{'}), 
\end{equation}
where $H$ and $H^{'}$ are the numbers of hyperons in mother and daughter 
nuclei, respectively. 
The hyper energy decreases also when the normal particles 
are emitted, In this case the additional hyper barrier is calculated 
with the same equation~(\ref{sep}) by taking $H^{'}=H$. 
We have included the ground and the excited states of 
the hyper particles with their masses 
taken from the experimental tables \cite{Ban90,Has06}. 
As in the case of the emission of normal particles \cite{Botvina87} their 
masses explicitly enter the calculations of $B_{j}$. 
Presently we are 
interested in emission from the excited hyper residue containing one (or 
maximum two) absorbed hyperon. As follows from dynamical calculations 
\cite{Bot11}, the capture of large numbers of hyperons is associated with 
more intensive collisions, which lead to higher excitation of residual 
nuclei; therefore, it comes up into another decay mode, 
e.g., multifragmentation 
(see below). Because the hyperon fraction is negligible in comparison 
with the total number of nucleons, we do not expect a considerable 
modification of the hyper nuclear properties as compared to the normal 
nuclear ones. In this case the level densities are taken as in the case of 
normal nuclei with the same mass number $A$. The inverse cross section is also 
taken as for reactions with normal nuclei by considering a neutron instead of 
a hyperon. We believe that all these approximations 
are sufficient for the first estimate of the evaporation of hyperons and 
light hyper clusters. 
However, it should be improved after obtaining more reliable data on 
hyperon-nucleon interaction and after the development of the advanced 
theoretical parametrizations. For the beginning we do not include 
the larger hyper particles with $A>6$ 
for the emission in the model, because their 
probability will be essentially lower. However, they (and their excited 
states) can be included within the described method too.

An important process of de excitation of heavy nuclei ($A \goo 100$) is the 
fission of nuclei. 
This process competes with particle emission, and it can also be 
simulated with the Monte Carlo method at the each step of the 
evaporation-fission cascade. 
Following the Bohr-Wheeler statistical approach, we assume that 
the partial width for the normal compound nucleus fission is proportional 
to the level density at the saddle point $\rho_{sp}(E)$ \cite{Bon95},
\begin{equation} \label{eq:fis}
\Gamma_{f}=
\frac{1}{2\pi\rho_{AZ}(E_{AZ}^{*})}\int_{0}^{E_{AZ}^{*}-B_{f}}
\rho_{sp}(E_{AZ}^{*}-B_{f}-E)dE,
\end{equation}
where $B_{f}$ is the height of the fission barrier, which is determined by 
the Myers-Swiatecki prescription. For approximation of $\rho_{sp}$ we have 
used the results of the extensive analysis of nuclear fissility and 
$\Gamma_{n}$/$\Gamma_{f}$ branching ratios; see Ref.~\cite{Bon95} for details 
and references. 

Similar to the evaporation case, we consider hypernuclei with a small number 
of absorbed hyperons (for the beginning $H$=1); therefore, we do not expect 
that the 
level density properties and the fission mechanism will change essentially 
in comparison with normal nuclei. The modification should concern the terms 
depending on the mass formulas because heavy hypernuclei are more strongly 
bound. For this reason the fission barrier for hypernuclei will be higher than 
that of normal nuclei. As a first approximation we assume that a hyper barrier 
$B_{f}^{hyp}$ should be added in addition to the Myers-Swiatecki barriers. 
The final hyper energy release in the fission will be 
\begin{equation} \label{fishr}
E_{0}^{hyp} =  E_{b}^{\rm hyp}(A,H) -  E_{b}^{\rm hyp}(A^{'},H). 
\end{equation}
Here $A$ and $A^{'}$ are the mass numbers of the mother and daughter nuclei 
which contain a hyperon. However, because the barrier is determined 
in the saddle point, where nuclear fragments are not separated completely, 
$B_{f}^{hyp}$ should be smaller than $E_{0}^{hyp}$. From our experience 
in normal fission we expect that the deformation of the surface at this point 
may take around one-half of the final value. So we assume that 
$B_{f}^{hyp} = E_{0}^{hyp}/2$. This is a quite conservative estimate which 
leads to increasing the fission barrier on about 0.5 MeV for heavy nuclei. 
There are other theoretical studies of the hypernuclei deformation 
\cite{Min09} which tell us that the increasing the barrier, e.g., for 
$^{238}_{\Lambda}U$ may vary approximately from 0.2 to 0.8 MeV. We emphasize 
that such an increase is very small compared to the excitation energy 
of the hyper residues: Usually $E_{AZ}^{*}$ is more than 10--20 MeV. This 
additional hyper barrier may slightly vary depending on masses of the 
formed fragments; however, the symmetric fission is the most probable at high 
excitations. Therefore, we have adopted $B_{f}^{hyp}$ at $A^{'}=A/2$ as 
a reasonable approximation for the calculation of the fission probability. 
We should take into account, of course, that for some specific nuclei at 
low excitation a more careful estimate of the fission barrier is necessary. 

The mass distribution of the produced fission fragments is calculated 
similar to normal fission events; see Ref.~\cite{Eren13}. Here we assume 
again that the small hyperon fraction cannot change the regularities 
established for normal nuclei, because the $\Lambda$-hyperon-nucleon 
interaction is qualitatively of the same order as nucleon-nucleon one. 
In this case a new uncertainty comes from the apparent deposition 
of the hyperon, in a bigger or a smaller fragment. We believe that 
it should be determined by the hyper binding energy, which is larger in a 
big fragment. The difference is 
\begin{equation} \label{duh}
\Delta U^{hyp} = E_{b}^{\rm hyp}(A1,H) -  E_{b}^{\rm hyp}(A2,H) , 
\end{equation}
where $A1$ and $A2$ are masses of these fragments ($A=A1+A2$). It is assumed 
that the probability $P1$ for the fragment $A1$ (if $A1>A2$) to get a hyperon 
can be found in the canonical way, 
\begin{equation} \label{pah}
P1 = 1/(1 + exp(- \Delta U^{hyp}/T)) , 
\end{equation}
where $T = \sqrt{E_{AZ}^{*} / (aA)}$ is the temperature of the system, 
and $a \approx 0.125$ is the level density parameter. Because the both 
$\Delta U^{hyp}$ and $T$ are of the order of MeV, this can lead to the 
essential redistribution of $\Lambda$-hyperons between big and small 
fragments. 

The kinetic energy of the hyper fission fragments is generated as for 
normal fragments \cite{Bon95,Eren13} and depends only on their mass number 
and charge. 
After fission the separated fragments are still excited and evaporate few 
particles. We use the above-described version of the evaporation for 
the calculation of this process. As a result we obtain the cold fissioning 
remnants and several free particles in the end of the Monte Carlo 
simulation of each event. 

All these models for secondary de excitation (evaporation and fission) 
were previously tested by numerical 
comparisons with experimental data on the decay of normal compound nuclei with 
excitation energies less than 2--3 MeV per nucleons 
\cite{Botvina87,Bon95,Eren13}. 
For this reason we expect that our extension of these models should give 
reliable predictions for the preparation of future experiments.

\subsection{Evolution from sequential decay to simultaneous break up}

The concept of the compound nucleus cannot be applied at high
excitation energies, $E^* \goo 3$ MeV/nucleon with the corresponding 
temperature $T\goo 5$ MeV. The reason is that the time
intervals between subsequent fragment emissions, estimated both within the
evaporation models \cite{charity} and from experimental data 
\cite{jandel,beau00,ISIS,Karn03}, 
become very short, on the order of a few tens of fm/$c$. In this case there will
be not enough time for the residual nucleus to reach equilibrium between
subsequent emissions.
Moreover, the produced fragments will be in the vicinity of each other and,
therefore, they should interact strongly. The rates of the particle emission
calculated as in the case of isolated compound nuclei will not be reliable
in this situation. There are many other theoretical arguments 
in favor of a simultaneous break up at high excitation energy. For 
example, the Hartree-Fock and Thomas-Fermi calculations predict that 
the compound nucleus will be unstable at high temperatures \cite{HFTF}. 
Sophisticated dynamical calculations have also shown that a nearly
simultaneous break up into many fragments is the only possible way for
the evolution of highly excited systems \cite{XXX}. 
There also exist several analyses of experimental data which reject the
binary decay mechanism of fragment production
via sequential evaporation from a compound nucleus at high excitation energy 
\cite{ISIS,hubele,Deses,napolit}. 
 
The picture of a simultaneous break up in some freeze-out volume is 
more justified at the high energy. Indeed, the time scales of less than 
100 fm/$c$ are extracted for multifragmentation reactions from experimental 
data \cite{beau00,Karn03}. There are many experimental and theoretical 
works demonstrating a smooth transition from evaporation and fission modes 
to the fast multifragmentation break up of the whole nuclei with 
increasing excitation energy above 3 MeV/nucleon 
\cite{Karn03,ISIS,aladin,dago99,Sch01,Bon95,Eren13}. 
This break up can be described 
by the statistical laws, in particular, the statistical multifragmentation 
model (SMM) \cite{Bon95}, where the disintegration channels are generated 
according to their statistical weight. 
The corresponding physics is related to the thermal expansion and density 
fluctuations of the nuclear matter. 
The SMM was previously generalized for hypernuclei in Ref.~\cite{Bot07}: 
The grand canonical approximation leads to the following average yields of 
individual fragments: 
\begin{eqnarray} \label{yazh} 
Y_{\rm AZH}=g_{\rm AZH} V_f\frac{A^{3/2}}{\lambda_T^3} 
{\rm exp}\left[-\frac{1}{T}\left(F_{AZH}(T,V)-\mu_{AZH}\right)\right]. 
\nonumber\\ 
\mu_{AZH}=A\mu+Z\nu+H\xi,~~~~
\end{eqnarray} 
Here $T$ is the temperature, $F_{AZH}(T)$ is the internal free energies of 
these fragments, $V_f$ 
is the free volume available for the translation motion of the fragments 
in the freeze-out $V$, 
$g_{\rm AZH}$ is the ground-state degeneracy factor of species 
$(A,Z,H)$, $\lambda_T=\left(2\pi\hbar^2/m_NT\right)^{1/2}$ is the 
nucleon thermal wavelength, and $m_N$ is the average 
nucleon mass. The chemical potentials $\mu$, $\nu$, and $\xi$ are 
responsible for the mass (baryon) number, charge, and strangeness 
conservation in the system. A transition from the compound hyper-nucleus to 
multifragmentation regime has already been demonstrated \cite{Bot07,Buy13}. The 
combination of all de-excitation modes in a universal hypernuclear model (as 
it was done in SMM \cite{Bon95}) will be a subject of our forthcoming works.



\section{Discussion of the results}

In the beginning it is important to demonstrate that different de-excitation 
mechanisms (evaporation, fission and multifragmentation) are connected 
with each other in the real disintegration process. In Figs.~1 and 2 we show 
how the mass distributions of produced fragments evolve with the excitation 
energy of the $^{209}$Bi source. These calculations are performed only for 
normal nuclei. One can see that at low excitations (1 MeV per nucleon) 
we have standard evaporation and fission fragments. With increasing 
excitation energy the channels of multifragmentation decay responsible for 
producing intermediate mass fragments come into force. Already at 2.5 
MeV per nucleon a considerable part of fragments with A$>$15 are 
produced in the fast multifragmentation break up. As was mentioned, 
we expect qualitatively the same evolution for hypernuclear systems too.

\begin{figure}[tbh]
\includegraphics[width=0.6\textwidth]{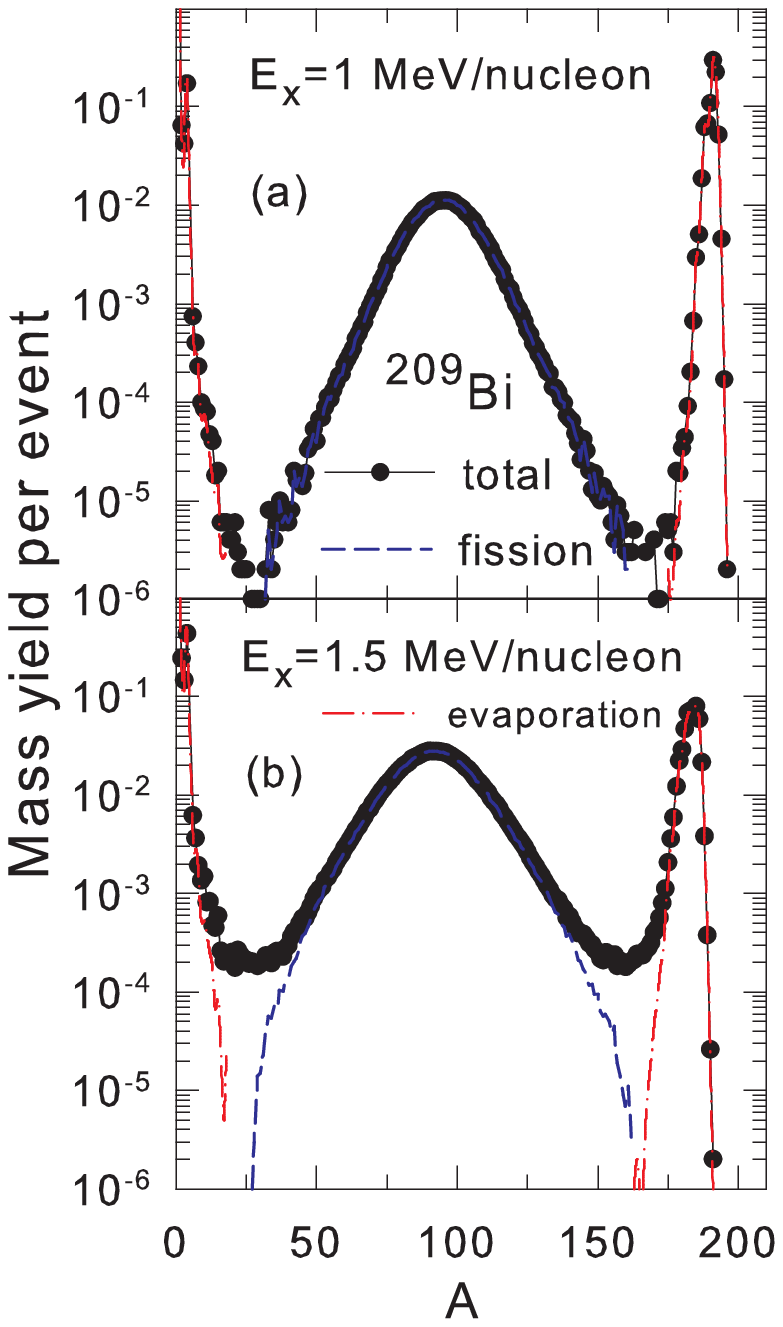}
\caption{\small{ (Color online)
Fragment mass distributions produced after disintegration of $^{209}$Bi 
excited normal nuclear systems. Excitation energies are given in panels (a), (b), and (c). The calculations include all decay processes for 
heavy nuclei: The contributions of fission (dashed blue lines) and 
evaporation (red dot-dashed lines) of the initial compound nuclei are shown 
separately. The rest to the total yields (solid lines with black circles) 
belong to the contribution of multifragmentation process. 
}}
\label{fig1}
\end{figure}

\begin{figure}[tbh]
\includegraphics[width=0.6\textwidth]{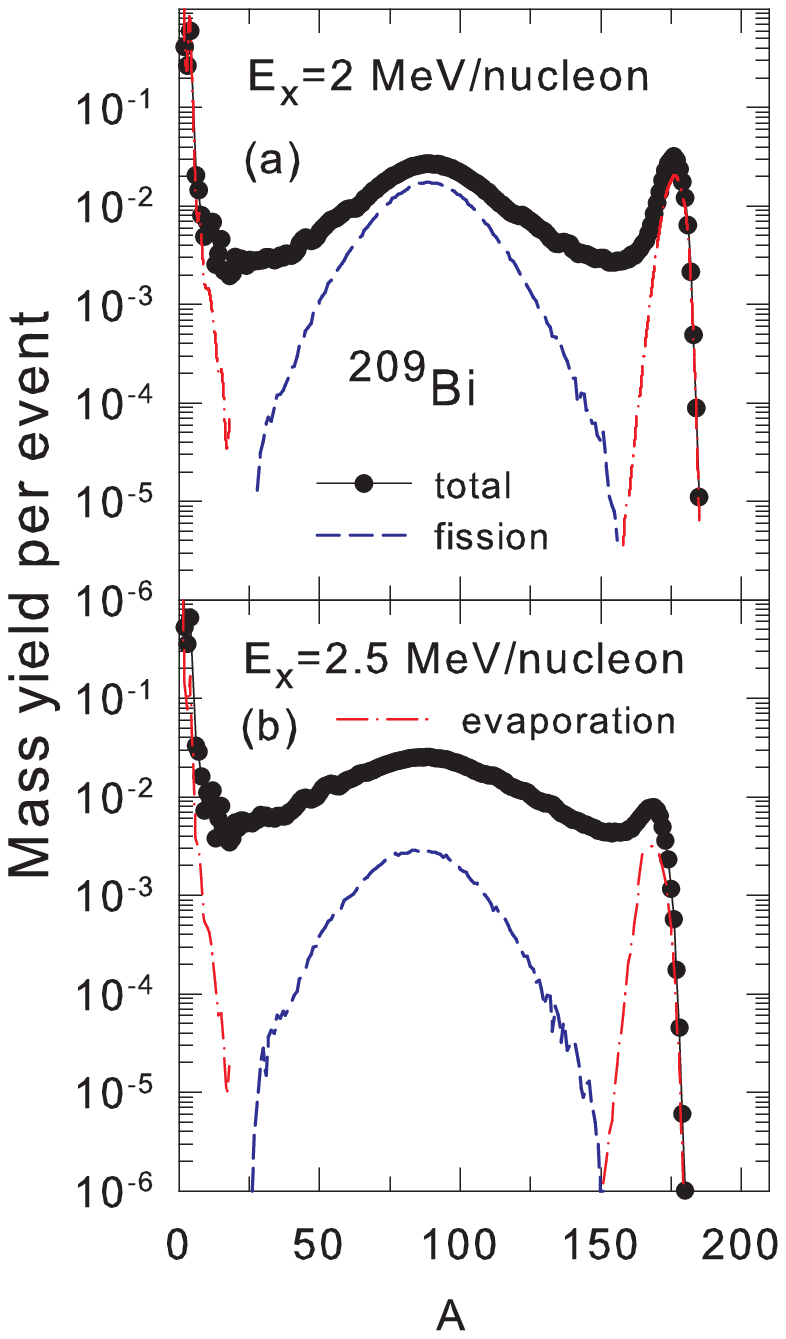}
\caption{\small{ (Color online) 
The same as in Fig.~1 but for higher 
excitation energies. 
}}
\label{fig2}
\end{figure}

The probabilities of the evaporation and fission processes can be 
easily measured in experiments. They are often used for testing the 
corresponding models. In Figs.~3 and 4 we demonstrate the 
evolution of these probabilities for heavy nuclei with their 
excitation energy. In particular, the "fission" means that the nuclei 
undergo fission during the de excitation at one of the steps of the 
evaporation-fission cascade. The label "evaporation residue" means that 
only evaporation of nucleons and light clusters take place during 
the de-excitation without fission. As one can see from calculations 
for normal nuclei (Fig.~3) the fission occurs certainly for very big 
nuclei ($^{238}$U) and it is strongly suppressed for medium-heavy nuclei 
($^{165}$Ho). In the same time the fission probability of medium-heavy 
nuclei increases with excitation energy essentially. Actually, it is 
consistent with the past experiments. For the big nuclei in between we 
can have obvious competition among these decay channels ($^{209}$Bi), and 
this can be measured in experiments too. We have shown this figure to facilitate the understanding the next Fig.~4, where the same 
processes are presented for evaporation and fission taking place in 
hyper nuclei.  

For all nuclear systems shown in Fig.~4 one $\Lambda$-hyperon is 
implemented in nuclei instead of one neutron. The single hypernuclei 
will be the most probable case in nuclear reaction, especially with 
deposition of low excitation energy \cite{BotG16}. The capture of many 
hyperons is usually accompanied by large nucleon loses, and, therefore, 
should lead to high excitation energies which are typical for 
multifragmentation channels \cite{Bot07,Ima15}. In Fig.~4 one can see 
similar trends for the fission and evaporation-residue probabilities 
as in Fig.~3. However, in this case one of the "hyper-fission" 
remnants will have a hyperon. Otherwise, the residue after 
evaporation (noted as "evap. $\Lambda$ --residue") can contain this $\Lambda$ hyperon. Another important channel shown 
in the Fig.~4 is the probability of the $\Lambda$-hyperon evaporation. 
One can see that this probability increases with excitation energy; 
however, it remains essentially smaller than the dominating process 
probability. It is because the $\Lambda$ binding energy keeps this hyperon 
inside the nuclei. 

\begin{figure}[tbh]
\includegraphics[width=0.5\textwidth]{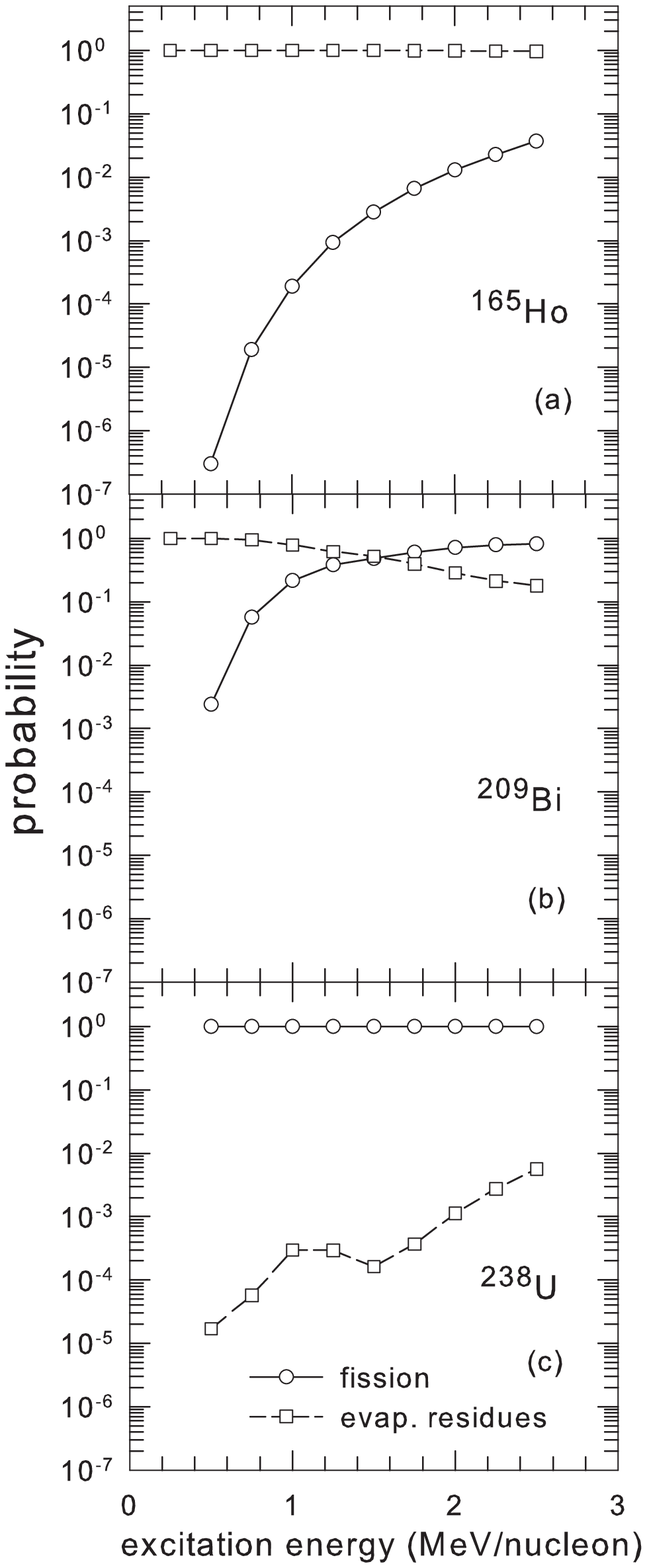}
\caption{\small{ 
Probability of the nuclear fission (circles, solid lines) and 
surviving the compound nucleus after evaporation of light 
particles (squares, dashed lines) versus excitation energy 
of nuclei. The nuclei are noted at the panels (a), (b), and (c). 
}}
\label{fig3}
\end{figure}

\begin{figure}[tbh]
\includegraphics[width=0.5\textwidth]{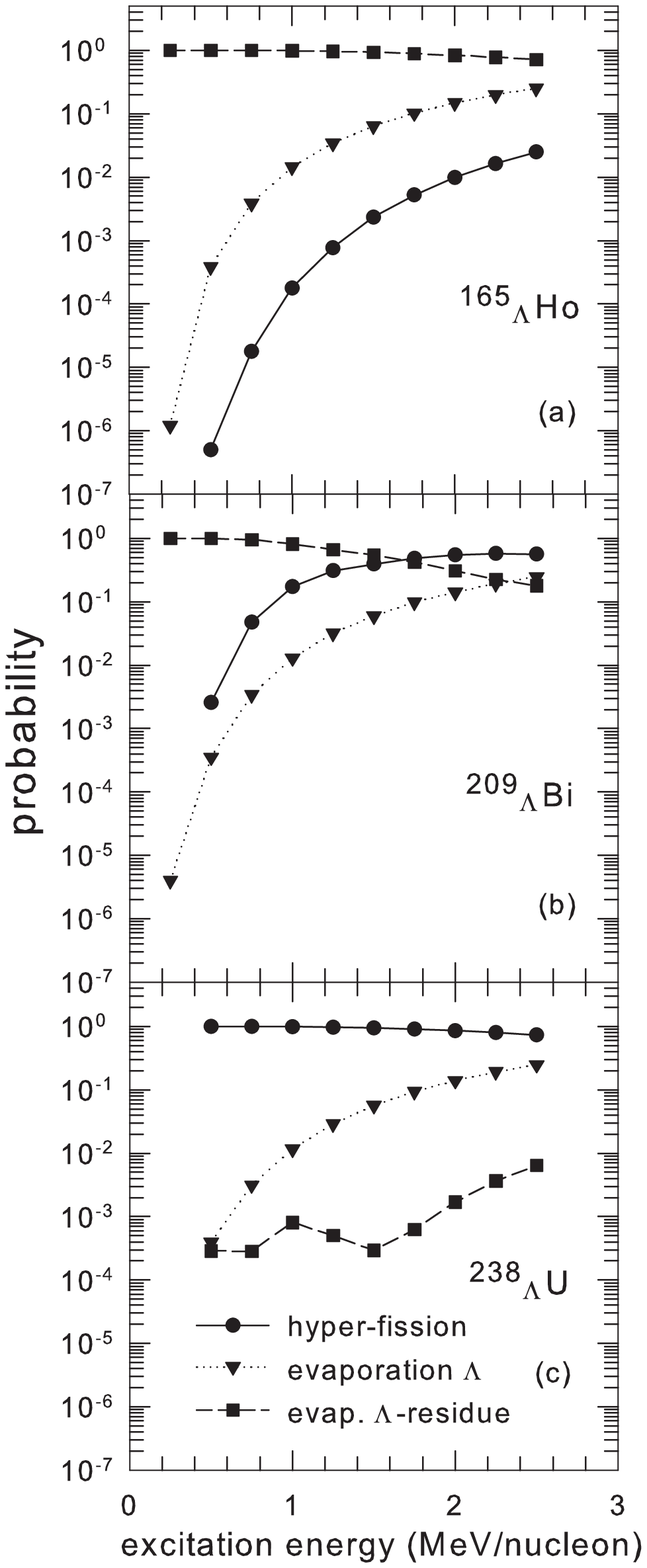}
\caption{\small{ 
Probability of the hyper-nuclear fission (solid circles, solid lines) 
and surviving the compound hyper-nucleus after evaporation of light 
particles (solid squares, dashed lines) versus excitation energy of 
hyper-nuclei. The solid triangles (dotted lines) give the probability 
for emission of single $\Lambda$-hyperons. 
The initial compound hyper-nuclei are noted in panels (a), (b), and (c). 
}}
\label{fig4}
\end{figure}

As was discussed above, besides the evaporation of $\Lambda$, 
an evaporation of light hyper-clusters is also possible from the excited 
hyper-compound. It is demonstrated in Fig.~5 for $^{3}_{\Lambda}$H, 
$^{4}_{\Lambda}$H, $^{4}_{\Lambda}$He, $^{5}_{\Lambda}$He, 
and $^{6}_{\Lambda}$He hyper-nuclei. Similar to the single 
$\Lambda$ evaporation the yield of these hyper nuclei increases 
considerably with increasing excitation energy. 
Still, the yield of these hypernuclei is essentially less than 
$\Lambda$-hyperons (more than one order of magnitude), and the 
probabilities of dominating evaporation/fission processes are much higher 
(compare Fig.~5 with Fig.~4). 
Evaporation of light clusters is well known in normal 
nuclear reactions. In the case of investigation of hypernuclei 
these processes may play a very important role, because they are 
produced as a result of a complex collective phenomena, but 
not as a result of a direct (or a coalescence-like) process in 
a final state. Previously, indications for existing of the 
unusual $\Lambda NN$ state coming from the disintegration of 
the excited projectile residues were reported \cite{Rap13}. 
They were never observed in direct reactions. This gives 
an opportunity to study how the formation of hypernuclei in new 
exotic states depends on the reaction mechanism. The lightest 
hypernuclei can be reliably identified in the projectile/target 
kinematic region by the decay correlation between the pions and 
normal fragments. As our calculations show (Fig.~5), the emission 
of $^{5}_{\Lambda}$He has largest probability, because of its 
considerable binding energy. As one can also see from the emulsion 
experiments \cite{Dav86}, these hypernuclei have the largest 
yield in comparison with others. 

\begin{figure}[tbh]
\includegraphics[width=0.5\textwidth]{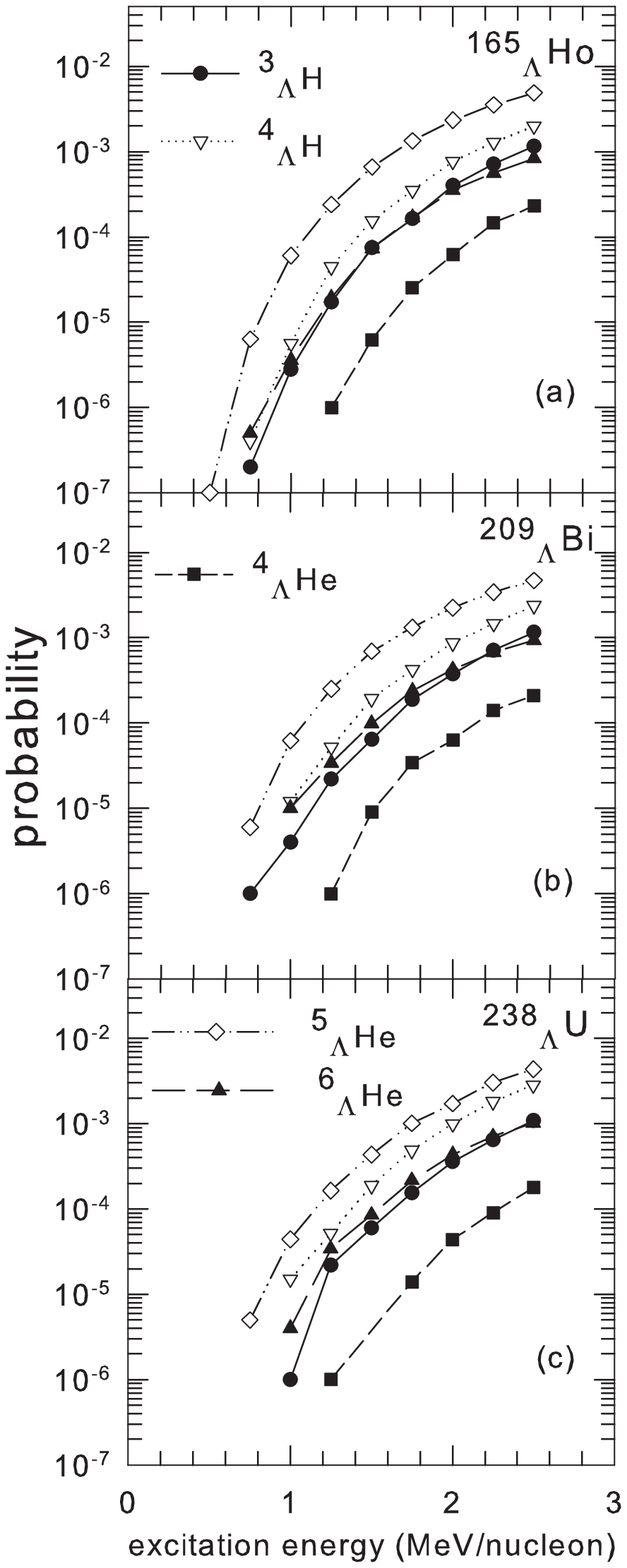}
\caption{\small{ 
Probability of the evaporation of light hyper-nuclei from 
heavy compound hyper-nuclei (see panels) versus excitation energy. The 
notations for evaporated nuclei (symbols and lines) are given 
in panels (a), (b), and (c).
}}
\label{fig5}
\end{figure}

Now we would like to draw more attention to the details of the 
hyper-fission process. As is well known, during the time of the 
deformation to the fission saddle point the compound nucleus can lose 
particles and the excitation energy via evaporation. Therefore, 
the nuclear composition at the saddle will be different than that 
in the beginning of the evaporation-fission cascade. We have plotted 
this effect in Figs.~6 and 7 depending on the excitation energy 
for the $^{209}_{\Lambda}$Bi case. The symbol "hot" corresponds to the 
parameters at the saddle point after which the fission process can become 
irreversible. The symbol "cold" presents  the average parameters of 
fission remnants after their scission and final evaporation cascade. 
The statistical deviations obtained in the Monte-Carlo 
simulations are shown by the 'error bars' separately above and below 
from the average values. 
We see that the initial nucleus can loose around ten nucleons during its 
evolution to the saddle deformation, and this pre-saddle emission 
increases with the 
excitation energy (Fig.~6). In addition the nucleus may lose a considerable 
part of the available energy (fig.~7). However, it is interesting that the 
fission barrier at the saddle can first decrease slightly at low excitations. 
This is because neutron are mainly evaporated, which increases the fissility. 
At high excitations, since many protons are evaporated also, the fissility 
decreases and the barrier becomes higher. Some residual excitation is 
predicted for the "cold" remnants and this excitation can be taken away by 
$\gamma$ emission. It may be used for examining hypernuclear structure too. 

\begin{figure}[tbh]
\includegraphics[width=0.6\textwidth]{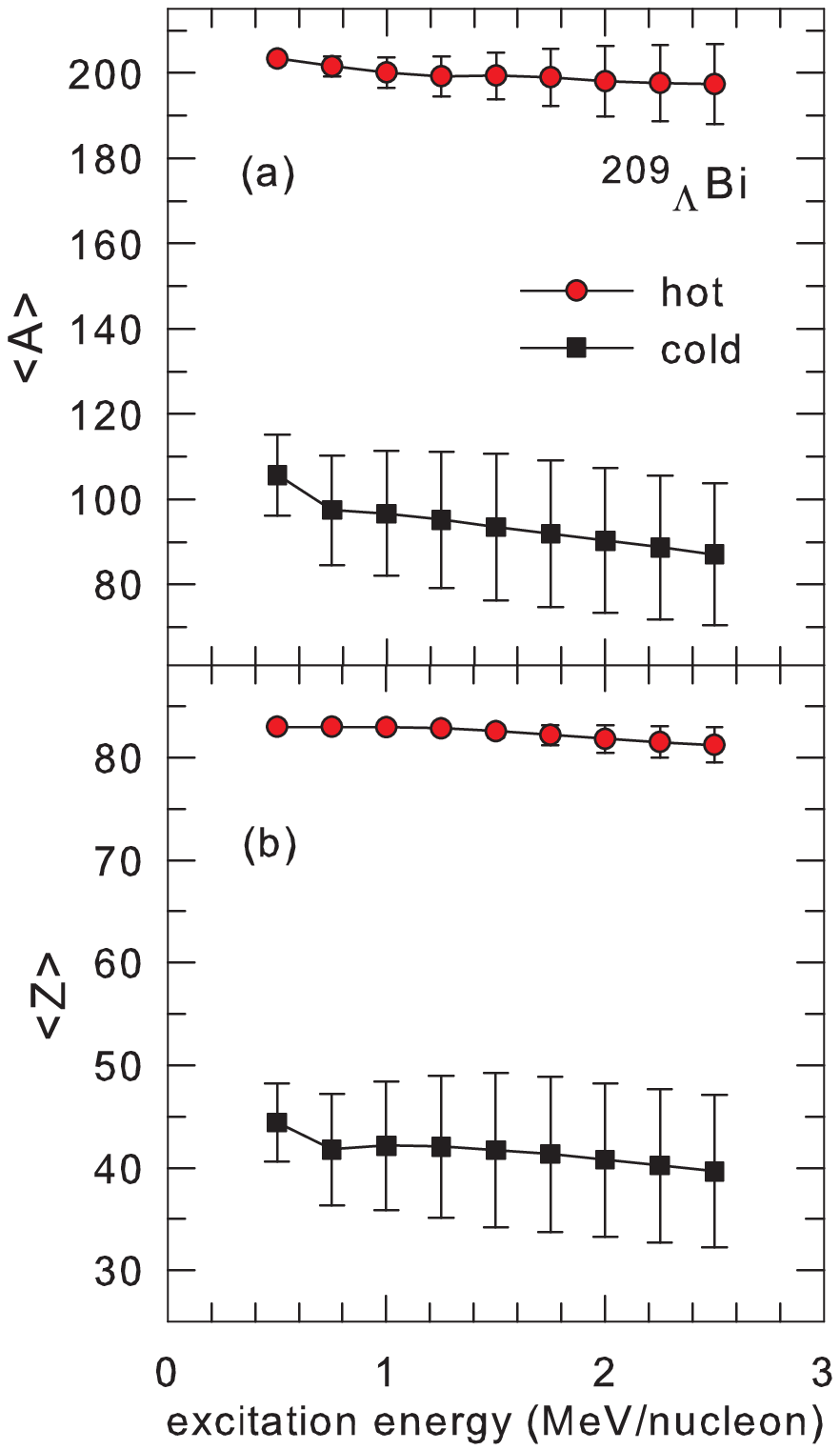}
\caption{\small{ (Color online)
Average mass numbers [top panel (a)] and charges [bottom panel (b)] 
for the fissioning $^{209}_{\Lambda}$Bi hyper-nuclei at the saddle 
point (noted as hot, red circles), and for final nuclear remnants 
after their de excitation (noted as cold, black squares), versus the 
excitation energy. The statistical deviations are shown by error bars. 
}}
\label{fig6}
\end{figure}

\begin{figure}[tbh]
\includegraphics[width=0.6\textwidth]{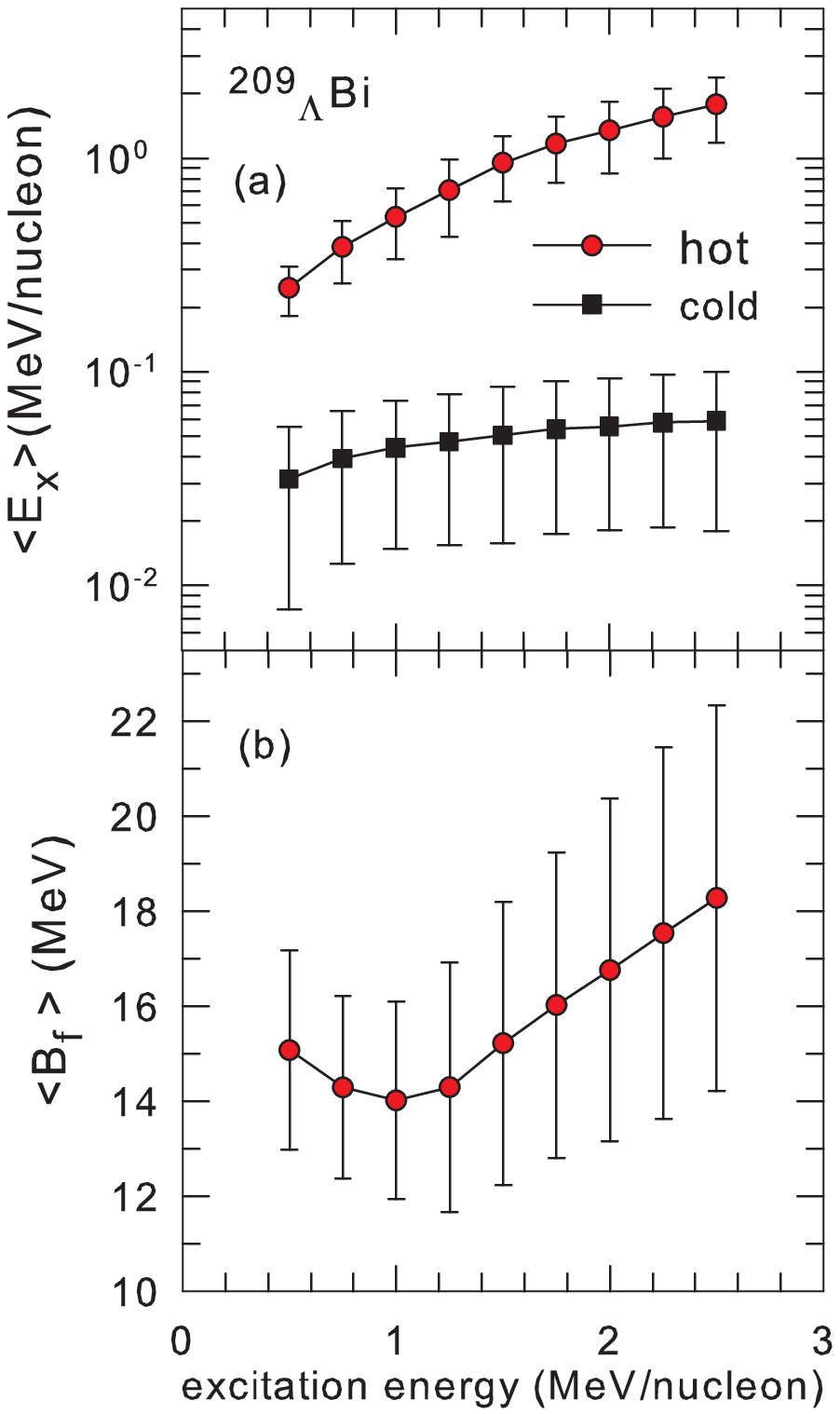}
\caption{\small{ (Color online)
The average excitation energy [$E^*$ (a)] and 
the fission barrier [$B_f$ (b)] 
of the fissioning $^{209}_{\Lambda}$Bi hyper nuclei at the saddle point 
(noted as hot) versus initial excitation energy. 
The remaining excitation energy of fission remnants is given on the panel (a) too. Other notations are as in Fig.~6.  
}}
\label{fig7}
\end{figure}

The typical characteristic for the fission process is the mass 
distribution of the fission fragments. In Figs.~8 and 9 we show the 
evolution of these mass distributions with excitation energy for 
fissioning $^{238}_{\Lambda}$U and $^{209}_{\Lambda}$Bi hypernuclei. 
As for the normal nuclear matter \cite{Bon95,Eren13} the uranium fission 
leads to the mixture of 
the asymmetric and symmetric modes at low excitation energy 
[Fig.~8(a)], which turn into the symmetric fission at 
more high excitations [Figs.~8(a) and (b)]. There are also 
small additions from the compound nucleus evaporation without 
fission. For the intermediate-heavy hypernuclei (Bi, Fig.~9) we expect 
only the symmetric fission with the considerable contribution from 
the compound nucleus evaporation. Both the evaporation and fission mass 
distributions become wider with the excitation. Note that in these figures 
we have presented the hyper-fragments only (i.e., they contain a 
$\Lambda$-hyperon). The complementary fission remnant is a normal fragment. 
Therefore, we can see in these figures slightly non-symmetric 
fission distributions, because the hyperons remain with greater probability 
in the largest fragments [see formula~(\ref{pah})]. It is especially 
obvious from panel (a) of Fig.~8, where the yield in the right 
asymmetric mode is higher than that in the left one. Previously, a similar 
distribution of hyperons between the remnants was reported in 
experiment \cite{Arm93}.

\begin{figure}[tbh]
\includegraphics[width=0.5\textwidth]{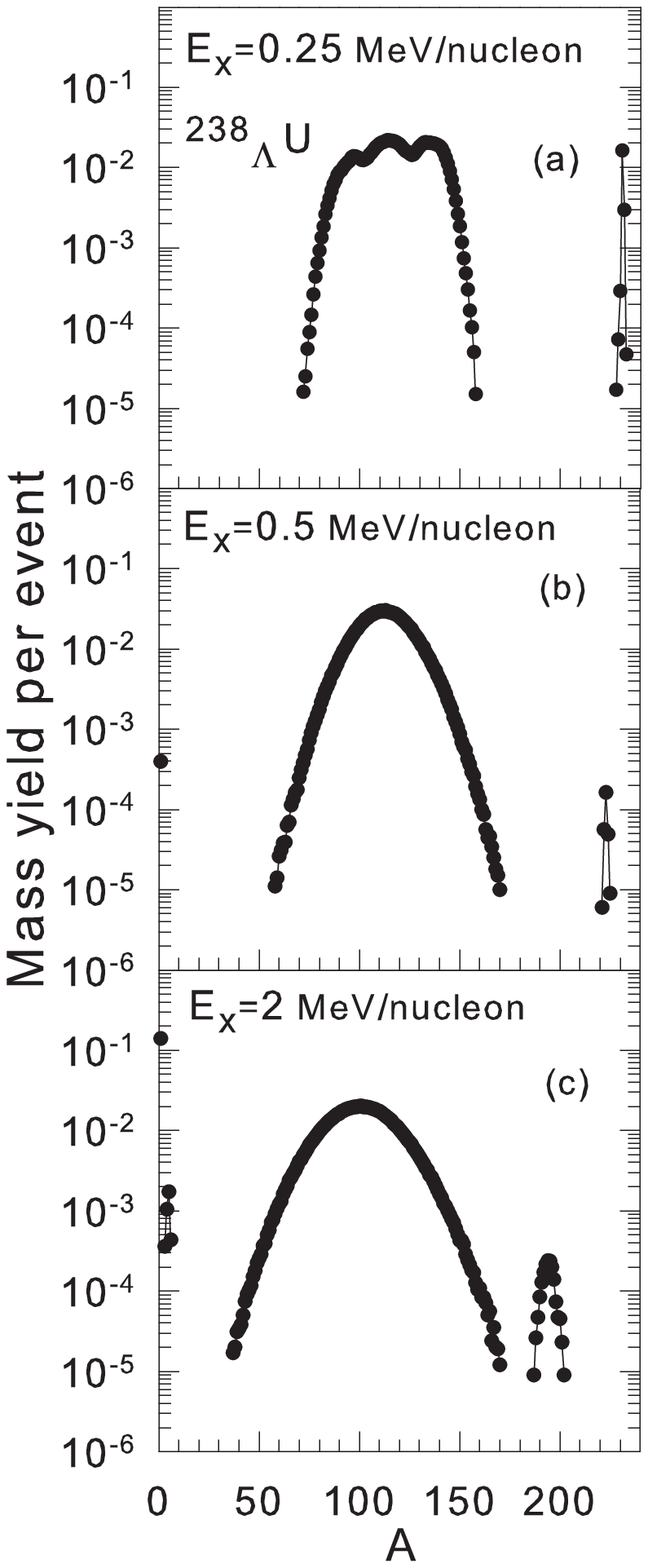}
\caption{\small{ 
Mass distributions of fission and evaporation hyper-fragments after 
disintegration of $^{238}_{\Lambda}$U hypernuclei at different 
excitation energies E$_{x}$ [in MeV per nucleon; see panels (a), (b), and (c)]. 
The calculations include the competition of evaporation and fission 
decay modes. 
}}
\label{fig8}
\end{figure}

\begin{figure}[tbh]
\includegraphics[width=0.5\textwidth]{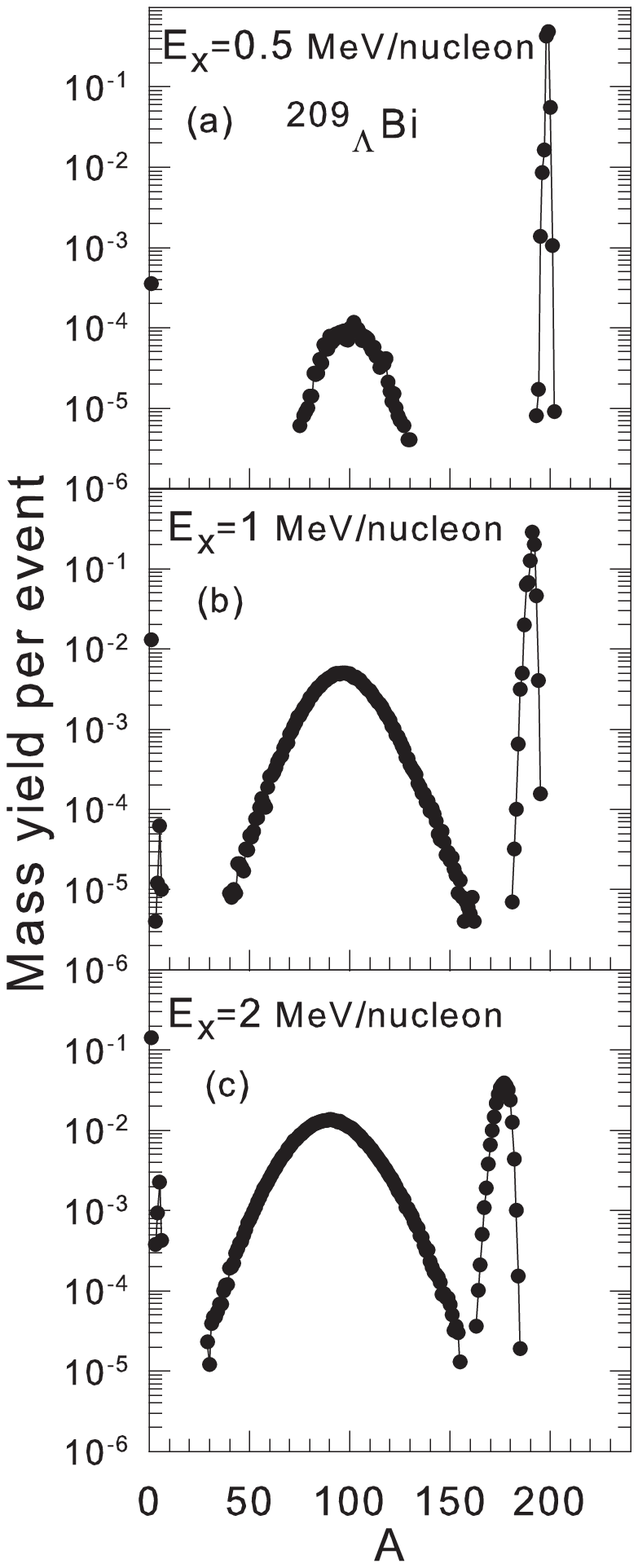}
\caption{\small{ 
The same as in Fig.~8, however, for $^{209}_{\Lambda}$Bi 
excited hypernuclei. 
}}
\label{fig9}
\end{figure}

Recently, the fission and evaporation processes in normal nuclei were 
discussed in the context of obtaining new nuclear isotopes. This is 
related to extending the nuclear chart and investigating the structure 
of exotic nuclei. We emphasize that involving hypernuclei provides 
novel opportunities for this research; see, e.g., discussion in our 
previous works \cite{Bot07,Buy13}. In Figs.~10--13 we demonstrate the 
isotope composition of hypernuclei produced as a results of 
evaporation-fission cascade. Actually, we calculate the probabilities 
of obtaining the isotopes in the case of de excitation initial U, Bi, and 
Sn hypernuclei at different excitation energies. The presentation of the 
results in the charge-neutron number plane is convenient for the 
overview and the selection of what reaction can be better for studying 
the specific isotopes. 

The results in Figs. 10 and 11 are obtained for two heavy 
fissioning nuclei, uranium and bismuth, which can capture a hyperon. 
Right above, at big charges close to the initial one,
we see the region of the compound hyper residues after the evaporation decay.
As expected, there are large neutron losses during the evaporation, which 
can be seen clear in comparison with the domain of the stable nuclei 
shown in the figures too. For the low excitation energy (0.25 MeV per 
nucleon) one can obtain a lot of intermediate neutron-rich hypernuclei 
as a result of the fission process. For normal nuclei, namely the fission 
reaction is considered as the promising method for obtaining the exotic 
neutron-rich nuclei. The presence of hyperons inside nuclei can 
increase their binding energy and we can get even a more exotic nuclear 
species \cite{Buy13}. With increasing excitation energy the subsequent 
de-excitation leads to neutron-poor hypernuclei, which would be also 
interesting for further studies. 
The general trends of the fragment yield after fissioning are qualitatively the same for both the heavy nuclei.

\begin{figure}[tbh]
\includegraphics[width=0.6\textwidth]{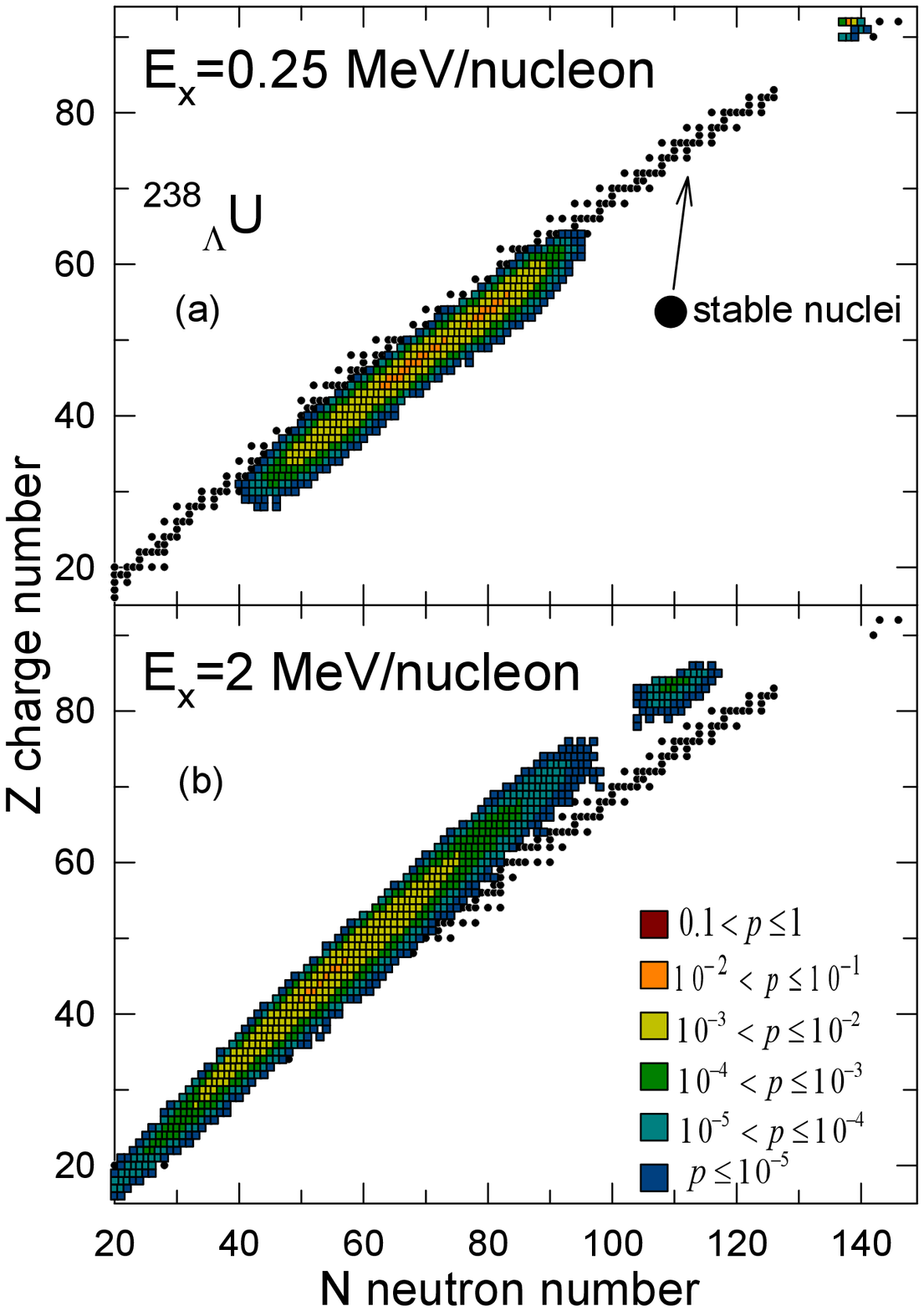}
\caption{\small{ (Color online)
Probabilities of the yield (normalized per one event) of single 
hypernuclei produced after evaporation and fission of the excited initial 
$^{238}_{\Lambda}$U hypernucleus. The squares present the nuclei in 
the plane of the charge number ($Z$) neutron number ($N$). The location of 
stable normal nuclei (from the nuclear chart) are represented by the dark 
circles to facilitate the comparison. 
Colors of the squares corresponding to the calculated ranges of the 
probability $p$ of these hypernuclei are given in the figure. 
Excitation energies are noted in panels (a) and (b). 
}}
\label{fig10}
\end{figure}

\begin{figure}[tbh]
\includegraphics[width=0.6\textwidth]{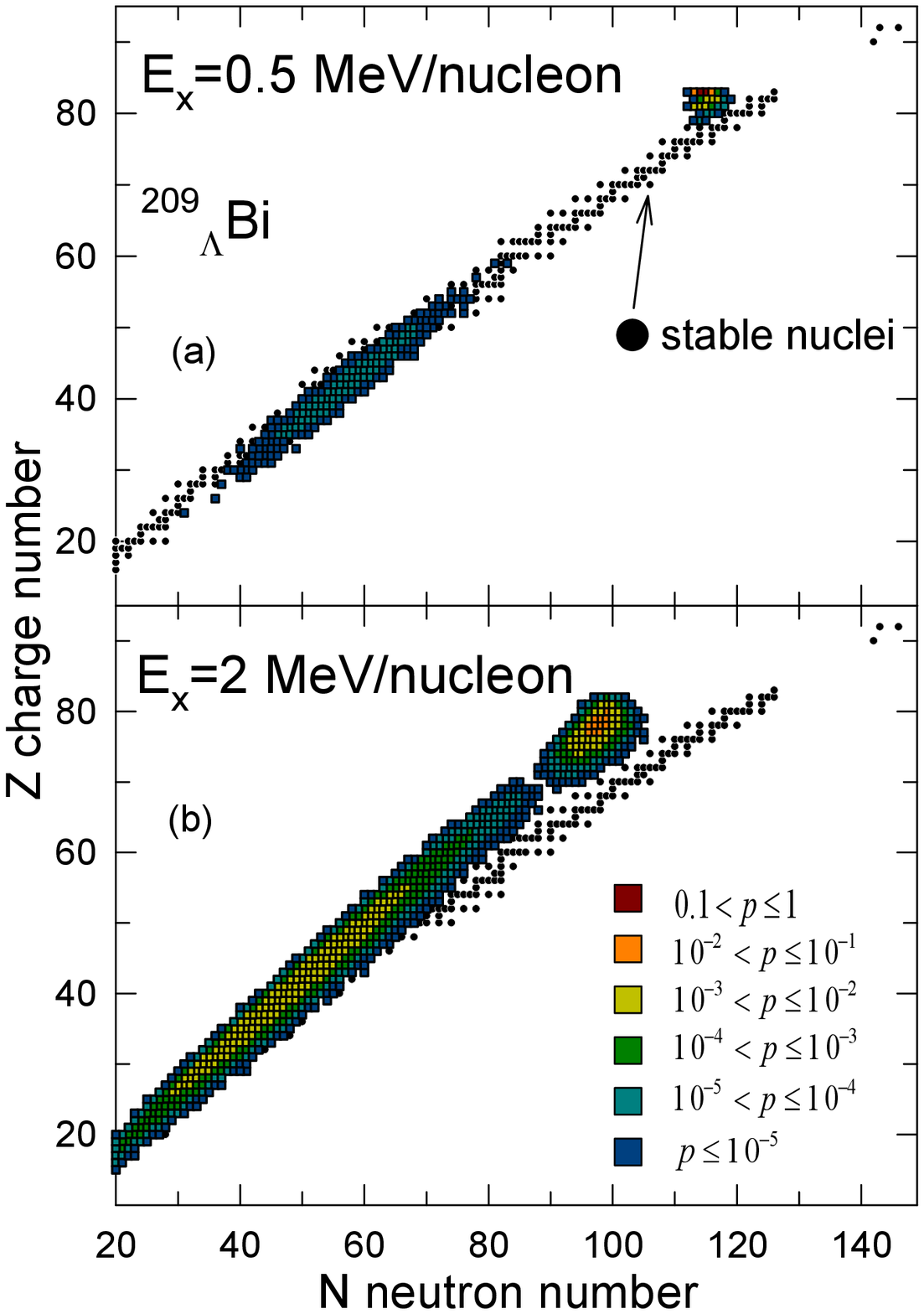}
\caption{\small{ (Color online)
The same as in Fig.~10, however, for the initial 
$^{209}_{\Lambda}$Bi hypernucleus. 
}}
\label{fig11}
\end{figure}

A promising nuclear method used in recent years is the comparative 
measurements for similar nuclei, however, with very different isospin.  
In our opinion, one of the important applications 
of newly obtained hypernuclear isotopes should be the 
studies of their lifetimes. It is known that the lifetime 
of $\Lambda$-hyperons inside nuclei is different from the lifetime of 
free $\Lambda$ \cite{Take14}. This is related to the subtle effects 
of weak interaction within nuclear matter. By producing neutron-rich 
and neutron-poor hypernuclei one can make a critical examination of the 
influence of nuclear isospin on the hyperon decay time. For this purpose 
one can use neutron-rich (-poor) projectiles (or targets) in relativistic 
ion collisions, that are possible to realize, e.g., at the GSI/FAIR facility 
\cite{aumann,frs}. For this reason in Figs.~12 and 13 we show the yield 
probabilities of hyper-elements coming after disintegration of tin 
hypernuclei with essentially different isospin content. By comparing 
$^{124}_{\Lambda}$Sn and $^{112}_{\Lambda}$Sn cases we see that the 
corresponding regions of final cold hypernuclei are well separated. 
In particular, the production of neutron-rich and neutron-poor hypernuclei 
does clear correlate with the initial isotope composition. 
In comparison with usual hypernuclear reactions, the procedure of the 
lifetime measurements is quite simple in the case of relativistic 
ions \cite{Take14}. Therefore, we think valuable information on the 
isospin dependence may be obtained rather soon in such experiments.

\begin{figure}[tbh]
\includegraphics[width=0.6\textwidth]{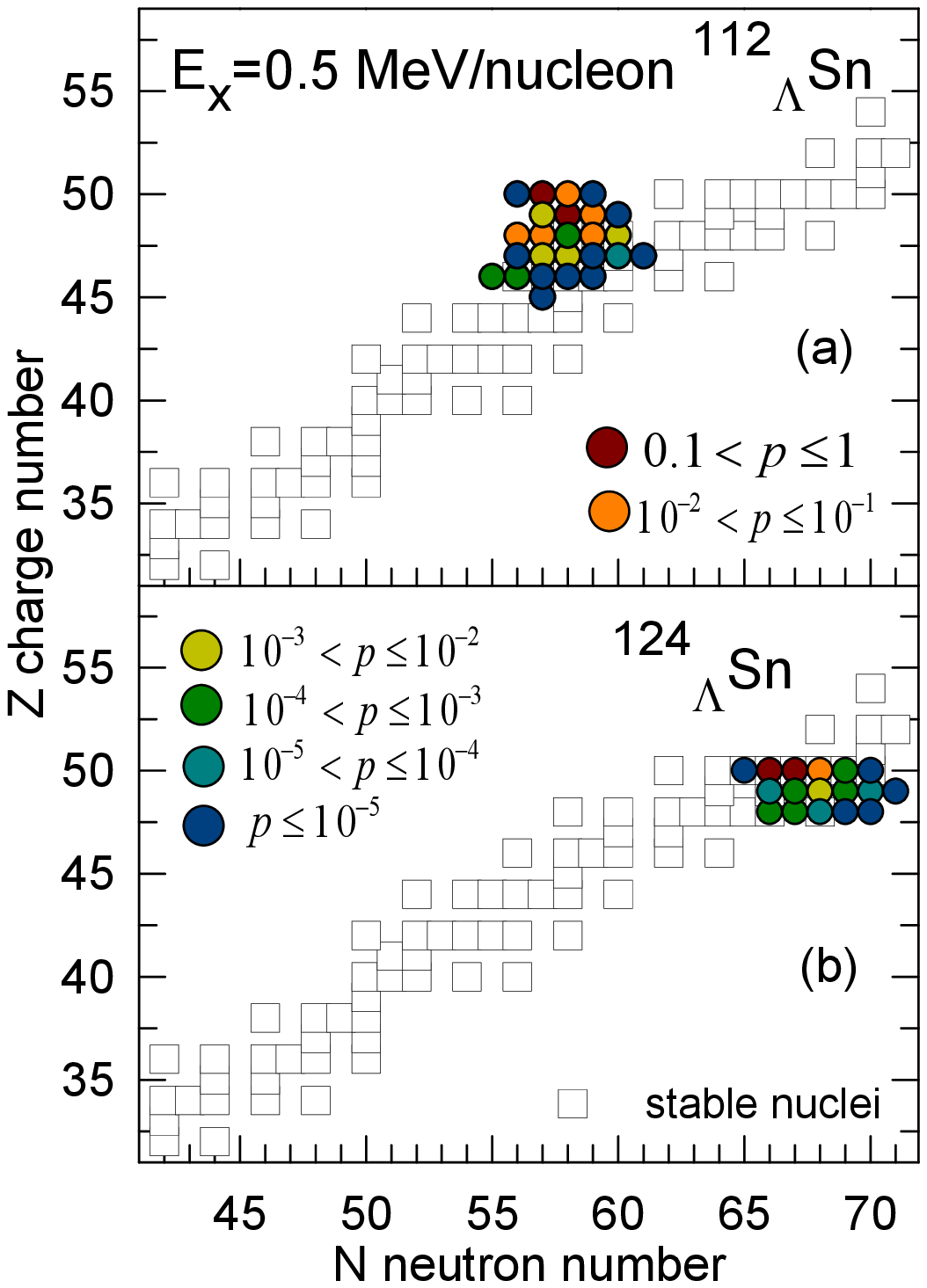}
\caption{\small{ (Color online)
Probabilities of the yield (normalized per one event) of single hypernuclei 
produced after evaporation and fission of the excited initial 
$^{112}_{\Lambda}$Sn (a) and $^{124}_{\Lambda}$Sn (b) hypernucleus at 
the excitation energy 0.5 MeV per nucleon. The circles present the nuclei in 
the plane of the charge number ($Z$) neutron number ($N$). The location of 
stable normal nuclei (from the nuclear chart) are represented by the empty 
squares. 
Colors of the circles corresponding to the calculated ranges of the 
probability $p$ of these hypernuclei are given in the figure. 
}}
\label{fig12}
\end{figure}

\begin{figure}[tbh]
\includegraphics[width=0.6\textwidth]{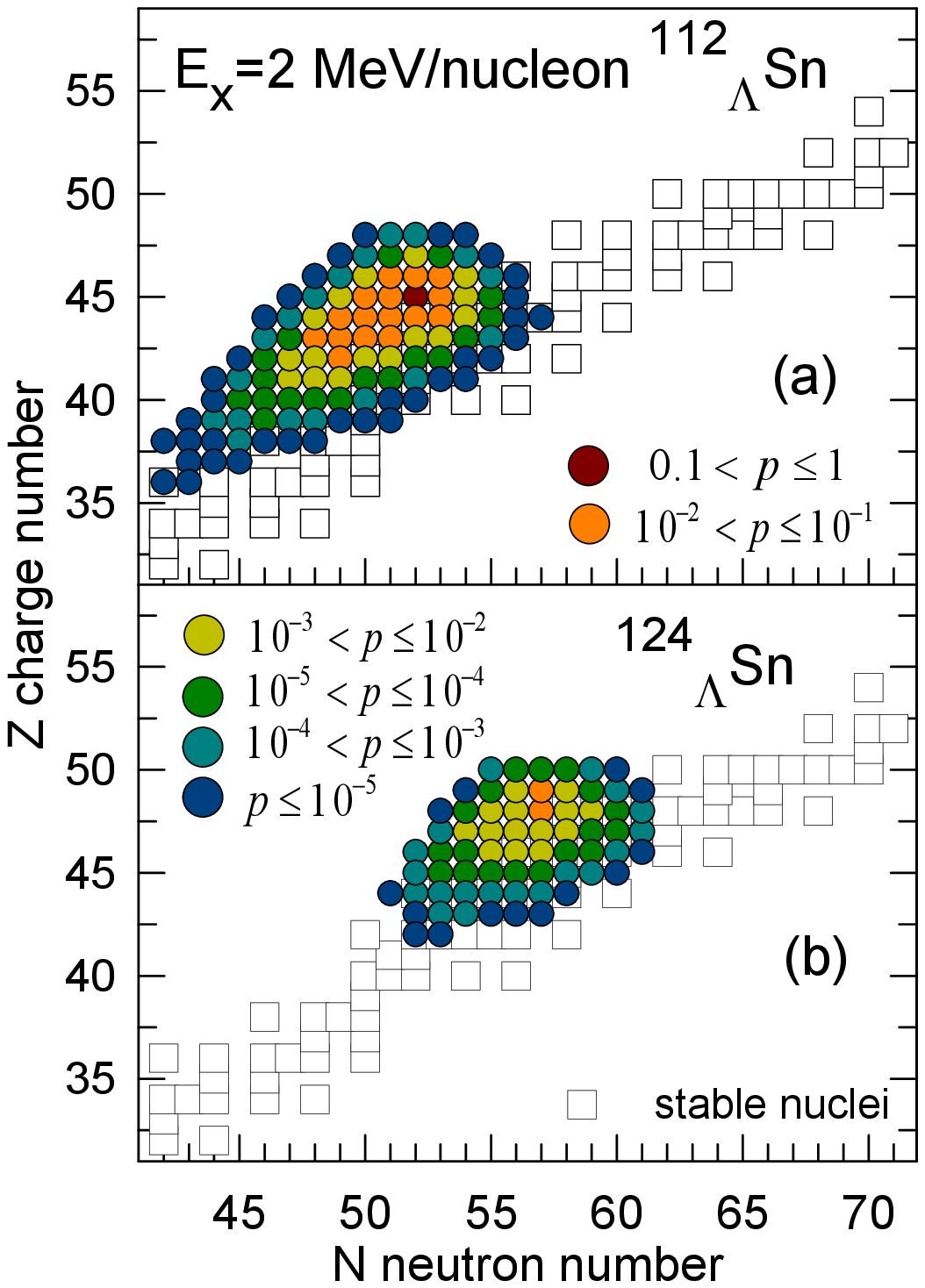}
\caption{\small{ (Color online)
The same as in Fig.~12, however, for the excitation energy 2 MeV per nucleon. 
}}
\label{fig13}
\end{figure}

\section{Conclusion}

The disintegration processes well established for excited normal nuclei 
can take place in hot hypernuclei also. It gives an opportunity to investigate 
the evolution of strange matter in such nuclei at relatively low temperature 
and obtain various hypernuclear states. 
Because the hyperon interaction within the matter is of the same order as 
the nucleon one, the extension of the reaction models designed for 
de excitation of normal nuclei becomes possible for  hypernuclei. 
In this work we have developed the models of the particle evaporation and 
the fission for excited hypernuclei. 
The critical ingredient which has to be included in the model 
is the binding energy of hyperons inside nuclei. 

We demonstrate that the results of hypernuclear evaporation and fission apparently taking place in deep inelastic nuclear 
collisions look similar to normal evaporation and fission processes. However, final 
hypernuclei obtained in this case offer a new direction for investigation. 
This may concern 
the production of exotic states which can exist because of the presence of 
a hyperon. It will be difficult to obtain such a state in other reactions, 
in particular, because there are practical limitations for using radioactive 
targets in experiments. We show the evaporational mechanism for emission 
of light hypernuclei which takes place from the target-projectile residues. 
Namely, the collective processes are responsible for formation of these small 
and large clusters. Therefore, contrary to phenomena of final state 
interactions for coalescence like and direct processes, novel 
hypernuclear states may be realized in this case. Important theoretical 
predictions are related to the fission process, which can be responsible 
for very neutron-rich hypernuclei. These nuclei can be used for many 
purposes, for example, by aiming at approaching the neutron-star conditions: 
There is a special interest in finding the weak decay lifetime dependence 
versus isospin. One can also extract the isospin 
dependence of the hyperon binding energy in neutron-rich matter 
via the comparison of the hypernuclei yields. 

The extension of the nuclear reaction research by involving captured 
hyperons will certainly have an impact on the field. It is a big 
advantage that in these reactions we obtain a very broad distribution of 
hypernuclei, as one can see from the calculated nuclear charts. 
Such nuclei can 
be immediately (on line) used for the extensive studies of their unknown 
properties. 

\begin{acknowledgments}

A.S. Botvina acknowledges the support of BMBF (Germany). 
N.B., A.E., and R.O. acknowledge the TUBITAK support under Project No. 114F328. N.B. and R.O. thank the Frankfurt 
Institute for Advanced Studies (FIAS), J.W. Goethe University for hospitality 
during the research visits. 

\end{acknowledgments}

\end{document}